\documentclass[12pt, a4paper, oneside]{article}

\newcommand{\ket}[1]{\left|#1\right\rangle}      
\newcommand{\bra}[1]{\left\langle #1\right|}     

\begin{document}

\newpage
\setcounter{page}{0}
\begin{titlepage}
\begin{flushright}
UFSCARF-TH-08-01
\end{flushright}
\vskip 1.0cm
\begin{center}
{\large  Functional relations from the Yang-Baxter algebra: \\
Eigenvalues of the $XXZ$ model with \\
non-diagonal twisted and open boundary conditions }\\
\vskip 1.5cm
{\large W. Galleas } \\
\vskip 1cm
{\em Universidade Federal de S\~ao Carlos\\
Departamento de F\'{\i}sica \\
C.P. 676, 13565-905, S\~ao Carlos-SP, Brasil}\\
\end{center}
\vskip 1.5cm

\begin{abstract}
In this work we consider a functional method in the theory of exactly solvable models
based on the Yang-Baxter algebra. Using this method we derive the eigenvalues of the $XXZ$ model
with non-diagonal twisted and open boundary conditions for general values of the anisotropy and
boundary parameters.
\end{abstract}

\vskip 2.5cm
\centerline{{\small PACS numbers:  05.50+q, 02.30.IK}}
\vskip 0.1cm
\centerline{{\small Keywords: Functional equations, Lattice Models, Open Boundary Conditions}}
\vskip 2.5cm
\centerline{{\small August 2007}}
\end{titlepage}

\section{Introduction}

Functional equations methods appeared in the theory of exactly solvable lattice models intimately connected
with Baxter's commuting transfer matrix method \cite{BAX0}. In the early seventies Baxter introduced in his 
pioneer work \cite{BAX0} the concept of $Q$-operators and $T-Q$ relations determining the eigenvalues of the 
transfer matrix of the corresponding vertex model.

The complex calculations involved in Baxter's construction of $Q$-operators seems to have restricted its use and other 
functional methods, such as the Reshetikhin's analytical Bethe ansatz, were employed instead to obtain the
spectrum of transfer matrices related with quantum Kac-Moody algebras \cite{RESH}. However we remark here  the recent
progresses in the construction of $Q$-operators by employing quantum algebras representation
theoretic methods \cite{ZAMO,KORFF}.

On the other hand, the advent of the algebraic Bethe ansatz in the late seventies provided a
systematic approach to find the eigenvalues and eigenvectors of transfer matrices of integrable vertex
models. This method, originally proposed by Takhtadzhan and Faddeev \cite{FAD0}, is based on the existence of a 
pseudovacuum or reference state and appropriate commutation rules provenient from the Yang-Baxter algebra,
which is a common algebraic structure associated with integrable vertex models.

Although the algebraic Bethe ansatz method is a powerful tool exhibiting a rich mathematical
structure, its implementation for models that do not possess a trivial reference state is still an
obstacle to be overcomed. The aim of this paper is to show that the Yang-Baxter algebra can be explored
in order to generate functional relations determining the spectrum of transfer matrices. In order to 
illustrate that we consider the $XXZ$ model with non-diagonal twists and open boundaries for general values
of the anisotropy and boundary parameters. Besides the relevance of studying models with general 
boundary conditions in the context of statistical mechanics \cite{ALC}, the $XXZ$ 
model with non-diagonal twists and open boundaries are both included in a class of models where a trivial
reference state is absent.

This paper is organized as follows. In section 2 we describe the $XXZ$ model with general toroidal
boundary conditions. In particular, we discuss the case of non-diagonal twists and we derive functional
relations for its eigenvalues making use of the Yang-Baxter algebra. In section 3 we approach the eigenvalue problem
for the $XXZ$ model with non-diagonal open boundaries using the algebraic-functional method devised
in the section 2. Concluding remarks are discussed in the section 4 and in the appendices A-D we give some
extra results and technical details.

\section{The $XXZ$ model with non-diagonal twisted boundary conditions}

The advent of the Quantum Inverse Scattering Method \cite{FAD0,KOR} was an important
stage in the development of the theory of exactly solvable quantum systems.
This method unveiled a deep connection between solutions of the Yang-Baxter
equation, quantum integrable systems and exactly solvable lattice models of
statistical mechanics in two dimensions \cite{BAX}. In statistical mechanics
an important role is played by vertex models whose respective
transfer matrix is constructed from local Boltzmann weights
contained in a operator $\mathcal{L}_{\mathcal{A} j}$.
Let $V$ be a finite dimensional linear space, the integrability
of the vertex model is achieved when the operator valued function
$\mathcal{L}: \mbox{\bf{C}} \rightarrow \mbox{End}( V \otimes V)$ is a solution of the
Yang-Baxter equation, namely
\begin{equation}
\label{ybe}
\mathcal{L}_{12}(\lambda - \mu) \mathcal{L}_{13}(\lambda) \mathcal{L}_{23}(\mu)
= \mathcal{L}_{23}(\mu) \mathcal{L}_{13}(\lambda) \mathcal{L}_{12}(\lambda - \mu),
\end{equation}
defined in the space $V_{1}\otimes V_{2}\otimes V_{3}$.
Here we use the standard notation $\mathcal{L}_{ij} \in \mbox{End}(V_{i} \otimes V_{j})$.
The complex valued operator $\mathcal{L}_{\mathcal{A} j}(\lambda)$ can be viewed as a matrix
in the space of states $\mathcal{A}$ denoting for instance the
horizontal degrees of freedom of a square lattice, while its matrix elements are operators acting
non-trivially in the $j$-th position of $\displaystyle \bigotimes_{i=1}^{L} V_{i}$. In its turn the
space $V_{j}$ represents the space of states of the vertical degrees of freedom at
the $j$-th site of a chain of lengh $L$.

The transfer matrix of the corresponding vertex model can be written in terms of the monodromy matrix
$\mathcal{T}_{\mathcal{A}}(\lambda)$ defined by the following ordered product
\begin{equation}
\label{mono}
\mathcal{T}_{\mathcal{A}}(\lambda) = \mathcal{L}_{\mathcal{A} L}(\lambda)
\mathcal{L}_{\mathcal{A} L-1}(\lambda) \dots \mathcal{L}_{\mathcal{A} 1}(\lambda).
\end{equation}
As a consequence of the Yang-Baxter equation, the monodromy matrix satisfies the following quadratic
relation
\begin{equation}
\label{yba}
R(\lambda - \mu) \; \mathcal{T}_{\mathcal{A}}(\lambda) \otimes \mathcal{T}_{\mathcal{A}}(\mu)
= \mathcal{T}_{\mathcal{A}}(\mu) \otimes \mathcal{T}_{\mathcal{A}}(\lambda) \; R(\lambda - \mu),
\end{equation}
usually denominated Yang-Baxter algebra. The $R$-matrix appearing in (\ref{yba}) follows
from the solution of the Yang-Baxter equation through the relation $R(\lambda)= P \mathcal{L}(\lambda)$
where $P$ denotes the usual permutation operator. The above $R$-matrix
plays the role of structure constant for the Yang-Baxter algebra and it consist of an invertible
complex valued matrix acting on the tensor product $\mathcal{A} \otimes \mathcal{A}$.

The invariance of the Yang-Baxter algebra plays an important role in the description of integrable
spin chains with general toroidal boundary conditions. One can easily verify the invariance of
(\ref{yba}) under the tranformation
$\mathcal{T}_{\mathcal{A}}(\lambda) \rightarrow G_{\mathcal{A}} \mathcal{T}_{\mathcal{A}}(\lambda)$
provided that the $c$-number matrix $G_{\mathcal{A}}$ is a symmetry of the $R$-matrix, i.e.
\begin{equation}
\label{rgg}
\left[ R(\lambda), G_{\mathcal{A}} \otimes G_{\mathcal{A}} \right] = 0 .
\end{equation}
Consequently we can define the operator
\begin{equation}
\label{trans}
T(\lambda) = \mbox{Tr}_{\mathcal{A}} \left[ G_{\mathcal{A}} \mathcal{T}_{\mathcal{A}}(\lambda) \right]
\end{equation}
which constitutes an one parameter family of commuting transfer matrices, i.e.
$\left[ T(\lambda), T(\mu) \right] = 0$.

Now restricting ourselves to the $XXZ$ model, $V_{i} \equiv \bf{C}^{2}$ and the corresponding
$\mathcal{L}$-operator is that of the anisotropic six vertex model
\begin{equation}
\label{loper}
\mathcal{L}(\lambda) = \pmatrix{
a(\lambda) & 0 & 0 & 0 \cr
0 & b(\lambda) & c(\lambda) & 0 \cr
0 & c(\lambda) & b(\lambda) & 0 \cr
0 & 0 & 0 & a(\lambda) \cr},
\end{equation}
whose Boltzmann weights are given by $a(\lambda)=\sinh(\lambda + \gamma)$,
$b(\lambda)=\sinh(\lambda)$ and $c(\lambda)=\sinh(\gamma)$.

In the Ref. \cite{BATCH1,BATCH2} the authors discuss the possible classes of twist matrices $G_{\mathcal{A}}$
compatible with the $R$-matrix associated with (\ref{loper}). These twist matrices
turn out to be
\begin{equation}
\label{gm}
(i) \;\;\; G_{\mathcal{A}} = \pmatrix{
\alpha & 0 \cr
0 & \beta \cr} \;\;\;\;\;\;\;\;\;\;\;\;\;\;\;\;\;\;
(ii) \;\;\; G_{\mathcal{A}} = \pmatrix{
0 & \alpha \cr
\beta & 0 \cr}
\end{equation}
where $\alpha$ and $\beta$ are arbitrary complex parameters. The
above matrices $G_{\mathcal{A}}$ are non-singular for $\alpha, \beta \neq 0$
and the $XXZ$ model with generalized toroidal boundary conditions is obtained as the
logarithmic derivative of the transfer matrix (\ref{trans}) at the point $\lambda=0$ \cite{BATCH1,BATCH2}.

The understanding of physical properties of vertex models demands the exact
diagonalization of their respective transfer matrices, which can provide us information about
the free energy behaviour and the nature of the elementary excitations.
When the twist matrix of type $(i)$ is considered,
a trivial reference state is available and the eigenvalues of the
corresponding transfer matrix can be obtained through the algebraic Bethe ansatz \cite{FAD0,KOR}.
By way of contrast, the twist matrix of type $(ii)$ breaks the $U(1)$ symmetry of the
the system leaving only a $Z_2$ invariance.

Although a trivial reference state in the framework of the algebraic Bethe ansatz is no longer
available when the matrix $G_{\mathcal{A}}$ of type $(ii)$ is considered,
the eigenvalues of the corresponding transfer
matrix were obtained in \cite{BATCH1} by means of the Baxter's $T-Q$ method.
It is worthwhile to remark here that the isotropic case
associated with the $XXX$ spin chain admits any $2\times 2$ twist
matrix and interesting enough the algebraic Bethe ansatz solution can be obtained
by exploring the $GL(2)$ symmetry \cite{GAP}.

From the historical point of view, the $T-Q$ method was introduced in Baxter's remarkable works on the
eight vertex model \cite{BAX0} and more recently it has found applications in many areas such as the study
of integrable systems \cite{refK1}, conformal field theory \cite{ZAMO,ZAMO12},
correlations functions \cite{KORF1} and efficient description of finite temperature properties \cite{KORF2}.
Motivated by the ideas of Baxter's $T-Q$ method \cite{BAX0,BAX} and the algebraic Bethe ansatz \cite{FAD0,KOR},
in what follows we shall demonstrate how we can use the Yang-Baxter algebra to obtain
functional relations determining the spectrum of the transfer matrix $T(\lambda)$.

Let us consider the eigenvalue problem for the transfer matrix
(\ref{trans}),
\begin{equation}
\label{eigp}
T(\lambda) \ket{\psi} = \Lambda(\lambda) \ket{\psi},
\end{equation}
taking into account the type $(ii)$ $G_{\mathcal{A}}$ matrix given in (\ref{gm}). As
shown in the appendix A, we can set $\alpha=\beta=1$ without loss of generality for the
purposes of this paper. Now considering the definition (\ref{mono}), the monodromy matrix
$\mathcal{T}_{\mathcal{A}}(\lambda)$ consist of a $2 \times 2$ matrix whose elements are
operators that we denote
\begin{equation}
\label{monorep}
\mathcal{T}_{\mathcal{A}}(\lambda) =\pmatrix{
A(\lambda) & B(\lambda) \cr
C(\lambda) & D(\lambda) \cr}.
\end{equation}
Therefore, the transfer matrix (\ref{trans}) reads
\begin{equation}
\label{trii}
T(\lambda) = B(\lambda) + C(\lambda).
\end{equation}

In contrast to the case when $G_{\mathcal{A}}$ is diagonal \cite{VEGA}, the state $\ket{0}$
defined as
\begin{equation}
\label{vac}
\ket{0} = \bigotimes_{j=1}^{L} \pmatrix{
1 \cr
0 \cr}
\end{equation}
is not an eigenstate of the transfer matrix (\ref{trii}). Nevertheless, the state $\ket{0}$
is still of great utility. Considering the definition (\ref{mono}) together with
(\ref{loper}), the elements of $\mathcal{T}_{\mathcal{A}}(\lambda)$ satisfy the following relations
\begin{eqnarray}
\label{action}
A(\lambda) \ket{0} &=& a(\lambda)^{L} \ket{0} \;\;\;\;\;\;\; D(\lambda) \ket{0} = b(\lambda)^{L} \ket{0} \nonumber \\
B(\lambda) \ket{0} &=& \dagger \;\;\;\;\;\;\;\;\;\;\;\;\;\;\;\;\;\;\;\; C(\lambda) \ket{0} = 0 ,
\end{eqnarray}
where the symbol $\dagger$ stands for a non-null value.
Notice the relations (\ref{action}) imply in
\begin{equation}
\label{T0}
T(\lambda) \ket{0} = B(\lambda) \ket{0}
\end{equation}
and
\begin{equation}
\label{TB0}
T(\lambda) B(\lambda) \ket{0} =  B(\lambda) B(\lambda) \ket{0} + C(\lambda) B(\lambda) \ket{0}.
\end{equation}

Now we have reached a point of fundamental importance. The term $C(\lambda) B(\lambda) \ket{0}$ present
in the right hand side of (\ref{TB0}) can be evaluated with the help of the Yang-Baxter
algebra (\ref{yba}). In order to show that we collect the following commutation rule among the ones
encoded in the relation (\ref{yba}),
\begin{equation}
\label{CXBY}
C(\lambda) B(\mu) = B(\mu) C(\lambda) + \frac{c(\lambda-\mu)}{b(\lambda-\mu)}
\left[ A(\mu) D(\lambda) - A(\lambda) D(\mu) \right].
\end{equation}
The commutation rule (\ref{CXBY}), together with the relations (\ref{action}), results in the following identity
\begin{equation}
\label{CB1}
C(\lambda) B(\mu) \ket{0} = \frac{c(\lambda - \mu)}{b(\lambda - \mu)}
\left[ a(\mu)^{L} b(\lambda)^{L} - a(\lambda)^{L} b(\mu)^{L} \right] \ket{0}.
\end{equation}
Therefore, as we are interested in $C(\lambda) B(\lambda) \ket{0}$, the next step is to consider the limit
$\mu \rightarrow \lambda$ in the above relation which can be evaluated using L'Hopital's rule.
Thus we are left with
\begin{equation}
\label{CB}
C(\lambda) B(\lambda) \ket{0} = M(\lambda) \ket{0}
\end{equation}
where $M(\lambda) = L c(\lambda)^2 a(\lambda)^{L-1} b(\lambda)^{L-1}$.

At this stage we have already gathered the basic ingredients to obtain functional relations determining 
the eigenvalues of $T(\lambda)$. Operating with the dual eigenvector $\bra{\psi}$ on the left side of
Eqs. (\ref{T0}) and (\ref{TB0}) we are left with the relations
\begin{eqnarray}
\label{f0}
\Lambda(\lambda) F_{0} &=& F_{1}(\lambda)  \\
\label{f1}
\Lambda(\lambda) F_{1}(\lambda)  &=& F_{2}(\lambda) + M(\lambda) F_{0},
\end{eqnarray}
where $F_{0}=\left\langle \psi | 0 \right \rangle$, $F_{1}(\lambda)=\bra{\psi} B(\lambda) \ket{0}$ and
$F_{2}(\lambda)=\bra{\psi} B(\lambda) B(\lambda) \ket{0}$.

Here we remark one of the roles played by integrability in this approach. Since the transfer matrix
$T(\lambda)$ belongs to a commuting family, the eigenvectors $\ket{\psi}$ are independent of the spectral
parameter $\lambda$. Thus the dependence of the functions $F_{i}(\lambda)$ with $\lambda$ is solely
determined by the operator $B(\lambda)$. In the appendix C we have computed that dependence making use of
Eqs. (\ref{mono}), (\ref{loper}) and (\ref{monorep}).

From the Eq. (\ref{f0}) we can see that the functions $\Lambda(\lambda)$ and $F_{1}(\lambda)$
differ only by the constant factor $F_{0}$, thus they possess the same zeroes.
Now we can eliminate $\Lambda(\lambda)$ from the Eq. (\ref{f0})
and substitute the result into the Eq. (\ref{f1}) which yields
the following relation involving only the functions $F_{i}(\lambda)$,
\begin{equation}
\label{f0f1}
F_{1}(\lambda)^2 = F_{2}(\lambda)F_{0} + M(\lambda) F_{0}^{2}.
\end{equation}

Considering the Eqs. (\ref{mono}), (\ref{loper}) and (\ref{monorep}), in the appendix C we have
demonstrated that the functions $F_{1}(\lambda)$ and $F_{2}(\lambda)$ can be written as
\begin{equation}
\label{a1a2}
F_{1}(\lambda) = F_{1}(0) \prod_{i=1}^{L-1} \frac{\sinh(\lambda_{i}^{(1)} -\lambda)}{\sinh(\lambda_{i}^{(1)})} \;\;\;\;\; \mbox{and} \;\;\;\;\;
F_{2}(\lambda) = F_{2}(0) \prod_{i=1}^{2(L-1)} \frac{\sinh(\lambda_{i}^{(2)} -\lambda)}{\sinh(\lambda_{i}^{(2)})} ,
\end{equation}
where $\lambda_{j}^{(i)}$ denote the zeroes of the function $F_{i}(\lambda)$.
Now a closer look in the Eqs. (\ref{f0}) and (\ref{f1}) reveals that we can obtain the eigenvalue
$\Lambda(\lambda)$ by determining the variables $\lambda_{j}^{(i)}$ together with the ratios
$\frac{F_{i}(0)}{F_{0}}$.

In order to obtain analogues of Bethe ansatz equations determining the variables $\lambda_{j}^{(i)}$,
we first consider the Eq. (\ref{f0f1}) at the points $\lambda = \lambda_{j}^{(1)}$. Then we find
\begin{equation}
\label{ba1}
F_{2}(\lambda_{j}^{(1)}) + M(\lambda_{j}^{(1)}) F_{0} = 0.
\end{equation}
Next we consider the points $\lambda = \lambda_{j}^{(2)}$ in Eq. (\ref{f0f1}) which yields
\begin{equation}
\label{ba2}
F_{1}(\lambda_{j}^{(2)})^2 - M(\lambda_{j}^{(2)}) F_{0}^{2} = 0.
\end{equation}
Furthermore, we observe by setting $\lambda=0$ in the Eqs. (\ref{f0}) and (\ref{f1}) that the ratios
$\frac{F_{i}(0)}{F_{0}}$ can be written in terms of $\Lambda(0)$ which can be easily evaluated.
Therefore we are left with the following expression for the ratios $\frac{F_{i}(0)}{F_{0}}$,
\begin{eqnarray}
\label{ratios}
\frac{F_{1}(0)}{F_{0}} &=&  \Lambda(0) \nonumber \\
\frac{F_{2}(0)}{F_{0}} &=&  \Lambda(0)^2 .
\end{eqnarray}
In order to avoid an overcrowded section we have performed the diagonalization of $T(0)$ in the appendix B.
The eigenvalue $\Lambda(0)$ turns out to be
\begin{equation}
\label{gama0}
\Lambda(0) = \sinh(\gamma)^L e^{\frac{i \pi r}{L}} \;\;\;\;\;\;\;\;\;\;\;\;\;\;\;\;\;\; r = 0, \dots, 2L-1.
\end{equation}

Gathering our results so far, from the Eqs. (\ref{f0},\ref{ba1},\ref{ba2},\ref{ratios}) we obtain
the following expression for the transfer matrix eigenvalues $\Lambda(\lambda)$,
\begin{equation}
\label{eigf}
\Lambda(\lambda) = \sinh(\gamma)^L e^{\frac{i \pi r}{L}} \prod_{i=1}^{L-1} \frac{\sinh(\lambda_{i}^{(1)}- \lambda)}{\sinh(\lambda_{i}^{(1)})}
\end{equation}
recalling that $r = 0, \dots, 2L-1$ and
provided that the variables $\lambda_{j}^{(i)}$ satisfy the following system of non-linear algebraic
equations
\begin{eqnarray}
\label{baf}
\left[ \frac{\sinh(\lambda_{j}^{(1)}+\gamma)}{\sinh(\gamma)} \frac{\sinh(\lambda_{j}^{(1)})}{\sinh(\gamma)} \right]^{L-1}
&=&-\frac{e^{\frac{2 i \pi r}{L}}}{L} \prod_{i=1}^{2(L-1)} \frac{\sinh(\lambda_{i}^{(2)}- \lambda_{j}^{(1)})}{\sinh(\lambda_{i}^{(2)})}
\;\;\;\;\;\;\;\; j=1,\dots, L-1  \nonumber \\
\\ \label{baf1}
\left[ \frac{\sinh(\lambda_{j}^{(2)}+\gamma)}{\sinh(\gamma)} \frac{\sinh(\lambda_{j}^{(2)})}{\sinh(\gamma)} \right]^{L-1}
&=&\frac{e^{\frac{2 i \pi r}{L}}}{L} \prod_{i=1}^{L-1} \left[ \frac{\sinh(\lambda_{i}^{(1)}- \lambda_{j}^{(2)})}{\sinh(\lambda_{i}^{(1)})} \right]^{2}
\;\;\;\;\;\;\;\; j=1,\dots, 2(L-1) \nonumber \\
\end{eqnarray}

Now we shall examine some aspects of the Eqs. (\ref{baf}) and (\ref{baf1})
taking into account the crossing properties discussed in the appendix D. Considering the Eq.
(\ref{DD}), it follows that
\begin{equation}
\label{gamat}
\Lambda^{t}(\lambda) = (-1)^{L+1} \Lambda(-\lambda-\gamma)
\end{equation}
where $\Lambda^{t}(\lambda)$ denotes the eigenvalue of the transposed transfer matrix $T^{t}(\lambda)$.
Although the relation (\ref{gamat}) does not imply in $\Lambda(\lambda) = (-1)^{L+1} \Lambda(-\lambda-\gamma)$,
we have verified the existence of eigenvalues satisfying that relation through
the direct diagonalization of $T(\lambda)$ for small chain lengh $L$.

Let us suppose the relation $\Lambda(\lambda) = (-1)^{L+1} \Lambda(-\lambda-\gamma)$ holds
for some eigenvalue. Hence,
from the Eq. (\ref{f0}) we can conclude that the function $F_{1}(\lambda)$ also satisfy
\begin{equation}
\label{y1}
F_{1}(\lambda) = (-1)^{L+1} F_{1}(-\lambda-\gamma) .
\end{equation}

Since the function $M(\lambda)$ enjoys the property
$M(\lambda) = M(-\lambda-\gamma)$, from the Eq. (\ref{f1}) we also find that
\begin{equation}
\label{y2}
F_{2}(\lambda) = F_{2}(-\lambda-\gamma) .
\end{equation}

Assuming that the relations (\ref{y1}) and (\ref{y2}) hold they have remarkable implications
concerning the solutions of Eqs. (\ref{baf}) and (\ref{baf1}).
Let us analyze first the odd $L$ case. Thus $L-1$ is an even number and
the Eq. (\ref{y1}) implies, for instance, in the following relation for
the variables $\lambda_{j}^{(1)}$,
\begin{equation}
\lambda_{i+\frac{L-1}{2}}^{(1)}= - \lambda_{i}^{(1)} - \gamma  \;\;\;\;\;\;\;\;\;\;\;\;\;\; i=1,\dots , \frac{L-1}{2} .
\end{equation}
Thus altogether we have only $\frac{L-1}{2}$ independent variables $\lambda_{j}^{(1)}$.

Now we turn our attention to the $L$ even case. In that case the Eq. (\ref{y1}) implies
for instance in the relation
\begin{equation}
\lambda_{i+\frac{L-2}{2}}^{(1)}= - \lambda_{i}^{(1)} - \gamma  \;\;\;\;\;\;\;\;\;\;\;\;\;\; i=1,\dots , \frac{L-2}{2} ,
\end{equation}
while the remaining root $\lambda_{L-1}^{(1)}$ is fixed at the value $-\frac{\gamma}{2}$.

The above analysis can also be carried out for the roots $\lambda_{j}^{(2)}$
taking into account the Eq. (\ref{y2}). Since $2(L-1)$ is an even number,
the variables $\lambda_{j}^{(2)}$ must be related for instance by
\begin{equation}
\lambda_{i+L-1}^{(2)}= - \lambda_{i}^{(2)} - \gamma \;\;\;\;\;\;\;\;\;\;\;\;\;\; i=1,\dots , L-1 .
\end{equation}

These implications of the crossing symmetry have been verified by a numerical
analysis of the Eqs. (\ref{baf}) and (\ref{baf1}) for small values of $L$. By comparing the eigenvalues
given by the Eq. (\ref{eigf}) with the direct diagonalization of
$T(\lambda)$, we have also verified that the Eqs. (\ref{eigf},\ref{baf},\ref{baf1}) indeed describes a complete spectrum.

We close this section by remarking the dependence with $\frac{1}{L}$ in the right hand side of
Eqs. (\ref{baf}) and (\ref{baf1}). As far as we know, this kind of dependence with the chain lengh
had not appeared previously in the context of Bethe ansatz, which may be of relevance for
the description of the thermodynamical limit $L \rightarrow \infty$.

\section{The $XXZ$ model with non-diagonal open boundaries}

In Sklyanin's pioneer work \cite{SK}, the author has been able to generalize the Quantum Inverse
Scattering Method to accomodate integrable spin chains with open boundaries. The
Yang-Baxter equation is still the corner stone of Sklyanin's approach and it turns out
that the $XXZ$ model with general open boundary conditions can be obtained from the
following double-row transfer matrix
\begin{equation}
\label{dtr}
t(u)= \mbox{Tr}_{\mathcal{A}} \left[ K^{+}_{\mathcal{A}}(u)
\mathcal{T}_{\mathcal{A}}(u) K^{-}_{\mathcal{A}}(u)
\bar{\mathcal{T}}_{\mathcal{A}}(u) \right]  ,
\end{equation}
where $\mathcal{T}_{\mathcal{A}}(u) = \mathcal{L}_{\mathcal{A} L}(u)
\mathcal{L}_{\mathcal{A} L-1}(u) \dots \mathcal{L}_{\mathcal{A} 1}(u)$
and $\bar{\mathcal{T}}_{\mathcal{A}}(u) = \mathcal{L}_{\mathcal{A} 1}(u)
\mathcal{L}_{\mathcal{A} 2}(u) \dots \mathcal{L}_{\mathcal{A} L}(u)$
are the standard monodromy matrices that generate the corresponding closed spin chain
with $L$ sites. The integrability at the boundaries is governed by the matrices
$K^{-}_{\mathcal{A}}(u)$ and $K^{+}_{\mathcal{A}}(u)$, each one describing the
reflection at one of the ends of an open chain.

Moreover, the boundary conditions compatible with the bulk integrability are constrained by the so-called reflection
equations, which for $K^{-}_{\mathcal{A}}(u)$ reads
\begin{equation}
\label{RE}
\mathcal{L}_{21}(u - v) K_{2}^{-}(u) \mathcal{L}_{12}(u + v) K_{1}^{-}(v) =
K_{1}^{-}(v) \mathcal{L}_{21}(u + v) K_{2}^{-}(u) \mathcal{L}_{12}(u - v),
\end{equation}
while a similar equation should also hold for the matrix $K^{+}_{\mathcal{A}}(u)$.

Turning our attention to the $XXZ$ model, the bulk hamiltonian is described
by the $\mathcal{L}$-operator (\ref{loper}) and the most general boundary
matrices satisfying the reflection equations are of the form
\begin{equation}
\label{KMat}
K_{\mathcal{A}}^{-}(u)=\pmatrix{
k^{-}_{11}(u) & k^{-}_{12}(u) \cr
k^{-}_{21}(u) & k^{-}_{22}(u) \cr}  \;\;\;\;\;\;
K_{\mathcal{A}}^{+}(u)=\pmatrix{
k^{+}_{11}(u) & k^{+}_{12}(u) \cr
k^{+}_{21}(u) & k^{+}_{22}(u) \cr},
\end{equation}
whose matrix elements are given by
\begin{eqnarray}
\label{kpw}
k^{-}_{11}(u) &=& \sinh(h^{-}_{1} + u) \;\;\;\;\;\;\;\;\;\;\;\;\;\;\;\;\;\;\;\;\; k^{-}_{12}(u) = h^{-}_{2} \sinh(2 u) \nonumber \\
k^{-}_{21}(u) &=& h^{-}_{3} \sinh(2 u) \;\;\;\;\;\;\;\;\;\;\;\;\;\;\;\;\;\;\;\;\;\;\; k^{-}_{22}(u) = \sinh(h^{-}_{1} - u)  \nonumber \\
k^{+}_{11}(u) &=& \sinh(h^{+}_{1} - u -\gamma) \;\;\;\;\;\;\;\;\;\;\;\;\;\;\; k^{+}_{12}(u) = h^{+}_{2} \sinh(-2 u -2\gamma) \nonumber \\
k^{+}_{21}(u) &=& h^{+}_{3} \sinh(-2 u - 2\gamma) \;\;\;\;\;\;\;\;\;\;\;\; k^{+}_{22}(u) = \sinh(h^{+}_{1} + u + \gamma).
\end{eqnarray}
The above $K$-matrices possess altogether six free boundary parameters $\{ h_{i}^{\pm} \}$
and we find the $XXZ$ model hamiltonian with open boundaries,
\begin{eqnarray}
\label{ham}
\mathcal{H} &=& \sum_{i=1}^{L-1} \sigma_{i}^{x} \sigma_{i+1}^{x} + \sigma_{i}^{y} \sigma_{i+1}^{y}
+ \cosh(\gamma) \sigma_{i}^{z} \sigma_{i+1}^{z} \nonumber \\
&+& \frac{\sinh(\gamma)}{\sinh(h_{1}^{-})} \left[ ( h_{2}^{-} + h_{3}^{-} )\sigma_{1}^{x} +
i( h_{2}^{-} - h_{3}^{-} )\sigma_{1}^{y} + \cosh(h_{1}^{-}) \sigma_{1}^{z} \right] \nonumber \\
&-& \frac{\sinh(\gamma)}{\sinh(h_{1}^{+})} \left[ ( h_{2}^{+} + h_{3}^{+} )\sigma_{L}^{x} +
i( h_{2}^{+} - h_{3}^{+} )\sigma_{L}^{y} + \cosh(h_{1}^{+}) \sigma_{L}^{z} \right],
\end{eqnarray}
related to the former by
\begin{equation}
\label{ft}
t'(0) = 2\sinh(h_{1}^{-}) \sinh(h_{1}^{+}) \sinh(\gamma)^{2L-1} \left[ \cosh(\gamma)
\mathcal{H} + L \cosh(\gamma)^2 + \sinh(\gamma)^2 \right],
\end{equation}
where $\sigma_{i}^{x}$, $\sigma_{i}^{y}$ and $\sigma_{i}^{z}$ denote
the usual Pauli matrices acting on the $i$-th site.

For general values of the boundary parameters $h_{i}^{\pm}$ the double-row transfer matrix
(\ref{dtr}) does not exhibit $U(1)$ invariance, which makes the application of the
algebraic Bethe ansatz quite complex due to the lack of a trivial reference state. 
By imposing $h_{2}^{\pm}=h_{3}^{\pm}=0$ the matrices $K_{\mathcal{A}}^{\pm}(\lambda)$ (\ref{KMat}) become
diagonal and the $U(1)$ symmetry exhibited by the bulk of the system is restored. In that case the spectrum of
(\ref{dtr}) was obtained in \cite{SK} through the algebraic Bethe ansatz. However, several progresses
concerning non-diagonal open boundaries have been
reported in the literature over the past years when the boundary parameters or the bulk anisotropy satisfy
a given constraint \cite{FAN}-\cite{WG1}.

In what follows we shall explore the Yang-Baxter algebra to generate functional relations determining
the complete spectrum of the double-row transfer matrix (\ref{dtr}) for general values of the bulk
anisotropy $\gamma$ and boundary parameters $h_{i}^{\pm}$. In order to tackle the eigenvalue
problem
\begin{equation}
t(u) \ket{\Phi} = \Delta(u) \ket{\Phi}
\end{equation}
for the double-row transfer matrix (\ref{dtr}),
we first notice the $\mathcal{L}$-operator (\ref{loper})
satisfies the property
$\mathcal{L}_{12}(u) \mathcal{L}_{12}(-u) = a(u) a(-u) \mbox{Id} \otimes \mbox{Id}$,
where $\mbox{Id}$ denotes the $2 \times 2$ identity matrix.
Therefore, the monodromy matrices $\mathcal{T}_{\mathcal{A}}(u)$ and
$\bar{\mathcal{T}}_{\mathcal{A}}(u)$ are related by
$\bar{\mathcal{T}}_{\mathcal{A}}(u) = \left[ a(u) a(-u) \right]^{L} \mathcal{T}^{-1}_{\mathcal{A}}(-u)$
and we find the following relation,
\begin{equation}
\label{rttinv}
\left[ \bar{\mathcal{T}}_{\mathcal{A}}(u) \otimes \mbox{Id} \right ]
R(2 u) \left[ \mathcal{T}_{\mathcal{A}}(u) \otimes \mbox{Id} \right]
= \left[ \mbox{Id} \otimes \mathcal{T}_{\mathcal{A}}(u) \right ] R(2u)
\left[ \mbox{Id} \otimes \bar{\mathcal{T}}_{\mathcal{A}}(u) \right ],
\end{equation}
with a straightforward manipulation of the Yang-Baxter algebra (\ref{yba}). Analogously to
Eq. (\ref{monorep}), we denote the matrix elements of $\bar{\mathcal{T}}_{\mathcal{A}}(u)$ by
\begin{equation}
\label{monobar}
\bar{\mathcal{T}}_{\mathcal{A}}(u) =\pmatrix{
\bar{A}(u) & \bar{B}(u) \cr
\bar{C}(u) & \bar{D}(u) \cr}.
\end{equation}
In this way the double-row transfer matrix (\ref{dtr}) reads
\begin{equation}
\label{dbr}
t(u) = k^{+}_{11}(u) U_{1}(u) + k^{+}_{22}(u) U_{2}(u) + k^{+}_{12}(u) U_{4}(u) + k^{+}_{21}(u) U_{3}(u)
\end{equation}
where
\begin{eqnarray}
\label{UU}
U_{1}(u) &=& k^{-}_{11}(u) A(u) \bar{A}(u) + k^{-}_{12}(u) A(u) \bar{C}(u) +
k^{-}_{21}(u) B(u) \bar{A}(u) + k^{-}_{22}(u) B(u) \bar{C}(u)  \nonumber \\
U_{2}(u) &=& k^{-}_{11}(u) C(u) \bar{B}(u) + k^{-}_{12}(u) C(u) \bar{D}(u) +
k^{-}_{21}(u) D(u) \bar{B}(u) + k^{-}_{22}(u) D(u) \bar{D}(u)  \nonumber \\
U_{3}(u) &=& k^{-}_{11}(u) A(u) \bar{B}(u) + k^{-}_{12}(u) A(u) \bar{D}(u) +
k^{-}_{21}(u) B(u) \bar{B}(u) + k^{-}_{22}(u) B(u) \bar{D}(u)  \nonumber \\
U_{4}(u) &=& k^{-}_{11}(u) C(u) \bar{A}(u) + k^{-}_{12}(u) C(u) \bar{C}(u) +
k^{-}_{21}(u) D(u) \bar{A}(u) + k^{-}_{22}(u) D(u) \bar{C}(u).  \nonumber \\
\end{eqnarray}
Next we consider the action of $t(u)$ on the state $\ket{0}$ defined in (\ref{vac}), which will require the
following relations
\begin{eqnarray}
\label{actiond}
\bar{A}(u) \ket{0} &=& a(u)^{L} \ket{0} \;\;\;\;\;\;\; \bar{D}(u) \ket{0} = b(u)^{L} \ket{0} \nonumber \\
\bar{B}(u) \ket{0} &=& \dagger \;\;\;\;\;\;\;\;\;\;\;\;\;\;\;\;\;\;\;\; \bar{C}(u) \ket{0} = 0
\end{eqnarray}
obtained from the structure of the monodromy matrix $\bar{\mathcal{T}}_{\mathcal{A}}(u)$,
and the relations (\ref{action}) from the previous section.

Then, in order to evaluate $t(u) \ket{0}$, we shall make use of the
commutation rules between the elements of $\mathcal{T}_{\mathcal{A}}(u)$ and
$\bar{\mathcal{T}}_{\mathcal{A}}(u)$ enclosed in the relation (\ref{rttinv}).
In particular, we shall use the following ones,
\begin{eqnarray}
\label{AAG}
A(u) \bar{B}(u) &=& \frac{b(2u)}{a(2u)} \bar{B}(u) A(u)
- \frac{c(2u)}{a(2u)} B(u) \bar{D}(u) \nonumber \\
D(u) \bar{B}(u) &=& \frac{a(2u)}{b(2u)} \bar{B}(u) D(u)
+ \frac{c(2u)}{b(2u)} \bar{A}(u) B(u) \nonumber \\
\bar{A}(u) B(u) &=& \frac{b(2u)}{a(2u)} B(u) \bar{A}(u)
- \frac{c(2u)}{a(2u)} \bar{B}(u) D(u) \nonumber \\
C(u) \bar{B}(u) &=& \frac{c(2u)}{a(2u)}
\left[ \bar{A}(u) A(u) - D(u) \bar{D}(u) \right] + \bar{B}(u) C(u) \nonumber \\
B(u) \bar{B}(u) &=& \bar{B}(u)  B(u).
\end{eqnarray}
Considering the Eqs. (\ref{dbr}), (\ref{UU}), (\ref{action}) and (\ref{actiond})
together with the relations (\ref{AAG}), a straightforward calculation
leave us with 
\begin{equation}
\label{tpsi0}
t(u) \ket{0} = Y_{0}(u) \ket{0} + Y_{1}(u) B(u) \ket{0} +
\bar{Y}_{1}(u) \bar{B}(u) \ket{0} + Y_{2}(u) B(u) \bar{B}(u) \ket{0},
\end{equation}
where the functions $Y_{0}(u), Y_{1}(u), \bar{Y}_{1}(u)$ and $Y_{2}(u)$ are given by
\begin{eqnarray}
\label{YI}
Y_{0}(u) &=& a^{2L}(u) \sinh( h_{1}^{+} - u ) \sinh( h_{1}^{-} + u )
\frac{\sinh(2u + 2\gamma)}{\sinh(2u + \gamma)}  \nonumber \\
&-& b^{2L}(u) \sinh(u + \gamma + h_{1}^{+}) \sinh(u + \gamma - h_{1}^{-})
\frac{\sinh(2u)}{\sinh(2u + \gamma)}  \nonumber \\
&-& a^{L}(u) b^{L}(u) \left( h_{2}^{-} h_{3}^{+} + h_{3}^{-} h_{2}^{+} \right)
\sinh(2u) \sinh(2u + 2\gamma) \nonumber \\
Y_{1}(u) &=&  a^{L}(u) h_{3}^{-} \sinh( h_{1}^{+} - u) \sinh(2 u)
\frac{\sinh(2u + 2\gamma)}{\sinh(2u + \gamma)} \nonumber \\
&+& b^{L}(u) h_{3}^{+} \sinh(u + \gamma - h_{1}^{-}) \sinh(2 u)
\frac{\sinh(2u + 2\gamma)}{\sinh(2u + \gamma)} \nonumber \\
\bar{Y}_{1}(u) &=& -a^{L}(u) h_{3}^{+} \sinh( h_{1}^{-} + u) \sinh(2 u)
\frac{\sinh(2u + 2\gamma)}{\sinh(2u + \gamma)} \nonumber \\
&+& b^{L}(u) h_{3}^{-} \sinh(u + \gamma + h_{1}^{+}) \sinh(2 u)
\frac{\sinh(2u + 2\gamma)}{\sinh(2u + \gamma)} \nonumber \\
Y_{2}(u)&=& -h_{3}^{+} h_{3}^{-} \sinh(2u) \sinh(2u + 2\gamma).
\end{eqnarray}
We can now follow the scheme devised in the previous section and operating with
the dual eigenvector $\bra{\Phi}$ on the left side of Eq. (\ref{tpsi0}) we obtain
\begin{equation}
\label{J0}
\Delta(u) f_{0}= Y_{0}(u) f_{0} + Y_{1}(u) f_{1}(u)
+ \bar{Y}_{1}(u) \bar{f}_{1}(u) + Y_{2}(u) f_{2}(u),
\end{equation}
where $f_{0}=\left\langle \Phi | 0 \right \rangle$, $f_{1}(u)=\bra{\Phi} B(u) \ket{0}$,
$\bar{f}_{1}(u)=\bra{\Phi} \bar{B}(u) \ket{0}$ and
$f_{2}(u)=\bra{\Phi} B(u) \bar{B}(u) \ket{0}$.

Before we proceed it is important to examine first the Eq. (\ref{J0}) under the light of the
results obtained in the appendix D.
Indeed, the crossing symmetry enables us to estabilish the relation
$B(u)=(-1)^{L+1} \bar{B}(-u-\gamma)$ (\ref{CCR}) which implies in
\begin{equation}
\label{ll1}
\bar{f}_{1}(u) = (-1)^{L+1} f_{1}(-u-\gamma).
\end{equation}
Furthermore, the commutation rule $\left[ B(u) , \bar{B}(u) \right] =0$ given in
(\ref{AAG}) together with the relation (\ref{CCR}), imply in the crossing invariance of the
function $f_{2}(u)$, i.e.
\begin{equation}
\label{ll2}
f_{2}(u) = f_{2}(-u-\gamma).
\end{equation}
From the results of the appendix D we also have the crossing invariance of the double-row
transfer matrix $t(u) =t(-u-\gamma)$, which implies in
\begin{equation}
\label{DC}
\Delta(u) = \Delta(-u-\gamma).
\end{equation}
In their turn the functions $Y_{i}(u)$ given in (\ref{YI}) fulfill the relations
$Y_{0}(u) = Y_{0}(-u-\gamma)$, $\bar{Y}_{1}(u) = (-1)^{L+1} Y_{1}(-u-\gamma)$ and
$Y_{2}(u) = Y_{2}(-u-\gamma)$. Taking into account the above properties, one can easily
verify the crossing invariance of the Eq. (\ref{J0}), which simplifies to the following
functional relation,
\begin{equation}
\label{JJ0}
\Delta(u) f_{0}= Y_{0}(u) f_{0} + Y_{1}(u) f_{1}(u)
+Y_{1}(-u-\gamma) f_{1}(-u-\gamma) + Y_{2}(u) f_{2}(u).
\end{equation}

Considering now the results from the appendix C, the function $f_{1}(u)$ (\ref{ww1}) can be written as
\begin{equation}
\label{m1}
f_{1}(u) = f_{1}(0) \prod_{i=1}^{L-1} \frac{\sinh(u_{i}^{(1)} - u)}{\sinh(u_{i}^{(1)})}.
\end{equation}
On the other hand, the functions $f_{2}(u)$ and $\Delta(u)$ given by (\ref{ww2}) and (\ref{w3}) will 
receive extra simplifications due to the relations (\ref{ll2}) and (\ref{DC}). Thus they can
be written as
\begin{eqnarray}
\label{m2}
f_{2}(u) &=& g_{2} \sinh(u) \sinh(u +\gamma) \prod_{i=1}^{L-2} \frac{\sinh(u_{i}^{(2)} - u)}{\sinh(u_{i}^{(2)})}
\frac{\sinh(u_{i}^{(2)} + \gamma + u)}{\sinh(u_{i}^{(2)}+\gamma)} \\
\Delta(u) &=& \Delta(0)  \prod_{i=1}^{L+2} \frac{ \sinh( u_{i}^{(0)} - u )}{\sinh(u_{i}^{(0)})}
\frac{ \sinh( u_{i}^{(0)} +\gamma + u )}{\sinh(u_{i}^{(0)}+\gamma)}.
\label{du}
\end{eqnarray}

The initial condition $\Delta(0)$ has been determined in the appendix B through a direct
analysis of $t(0)$. Then the next step is to determine the variables $u_{j}^{(i)}$.
In order to do that we consider the Eq. (\ref{JJ0}) at the points $u=u_{j}^{(0)}$, $u=u_{j}^{(1)}$ and $u=u_{j}^{(2)}$.
By doing so we find the following system of algebraic equations,
\begin{eqnarray}
\label{ui}
Y_{0}(u_{j}^{(0)}) f_{0} + Y_{1}(u_{j}^{(0)}) f_{1}(u_{j}^{(0)}) + Y_{1}(-u_{j}^{(0)}-\gamma) f_{1}(-u_{j}^{(0)}-\gamma) + Y_{2}(u_{j}^{(0)}) f_{2}(u_{j}^{(0)}) &=& 0 \\
\left[ Y_{0}(u_{j}^{(1)}) - \Delta(u_{j}^{(1)}) \right] f_{0} + Y_{1}(-u_{j}^{(1)}-\gamma) f_{1}(-u_{j}^{(1)}-\gamma) + Y_{2}(u_{j}^{(1)}) f_{2}(u_{j}^{(1)}) &=& 0 \\
\left[ Y_{0}(u_{j}^{(2)}) - \Delta(u_{j}^{(2)}) \right] f_{0} + Y_{1}(u_{j}^{(2)}) f_{1}(u_{j}^{(2)}) + Y_{1}(-u_{j}^{(2)}-\gamma) f_{1}(-u_{j}^{(2)}-\gamma) &=& 0.
\label{ui2}
\end{eqnarray}
A direct inspection of the Eqs. (\ref{ui}-\ref{ui2}) shows that the ratios
$\frac{f_{1}(0)}{f_0}$ and $\frac{g_{2}}{f_0}$ are also required. In contrast to the case of twisted boundary conditions
considered in the previous section, where $M(0)=0$, the point $u=0$ of Eq. (\ref{JJ0}) is not very
enlightening.
However, in order to determine the ratios $\frac{f_{1}(0)}{f_0}$ and $\frac{g_{2}}{f_0}$ we consider, for
instance, the non-trivial points $u=u_{1}$ and $u=u_{2}$ such that $Y_{1}(u_{1})=Y_{1}(u_{2})=0$. Then we are left with
the following equations
\begin{equation}
\label{rts}
\left[ Y_{0}(u_{i}) - \Delta(u_{i}) \right] f_{0} + Y_{1}(-u_{i} - \gamma) f_{1}(-u_{i} - \gamma)
+ Y_{2}(u_{i}) f_{2}(u_{i}) =0 \;\;\;\;\;\;\;\;\; i=1,2
\end{equation}
which can be solved for the required ratios.

Now our results can be gathered and we are left with the following expression for the eigenvalues
$\Delta(u)$,
\begin{equation}
\label{D1}
\Delta(u) = 2 \cosh(\gamma) \sinh(h_{1}^{+}) \sinh(h_{1}^{-}) \sinh(\gamma)^{2L}  \prod_{i=1}^{L+2} \frac{ \sinh( u_{i}^{(0)} - u )}{\sinh(u_{i}^{(0)})}
\frac{ \sinh( u_{i}^{(0)} +\gamma + u )}{\sinh(u_{i}^{(0)}+\gamma)},
\end{equation}
provided that the variables $u_{1}$, $u_{2}$ \footnote{We have performed the shifts
$u_{i} \rightarrow u_{i} - \frac{\gamma}{2}$.} and $u_{j}^{(i)}$ satisfy the following system
of non-linear algebraic equations,
\begin{equation}
\left[ \frac{\sinh(u_{i} + \frac{\gamma}{2})}{\sinh(u_{i} -\frac{\gamma}{2} )} \right]^L \frac{h_{3}^{-}}{h_{3}^{+}}
\frac{\sinh(u_{i} -\frac{\gamma}{2} - h_{1}^{+})}{\sinh(u_{i} + \frac{\gamma}{2} - h_{1}^{-})} = 1 
\;\;\;\;\;\;\;\;\;\;\;\;\;\;\;\; i=1,2
\end{equation}
 
\begin{eqnarray}
&&Y_{0}(u_{j}^{(0)}) = \nonumber \\
&&-\left[ Y_{1}(u_{j}^{(0)}) \prod_{i=1}^{L-1} \frac{\sinh(u_{i}^{(1)} - u_{j}^{(0)})}{\sinh(u_{i}^{(1)})}
+ Y_{1}(-u_{j}^{(0)}-\gamma) \prod_{i=1}^{L-1} \frac{\sinh(u_{i}^{(1)} + \gamma + u_{j}^{(0)})}{\sinh(u_{i}^{(1)})} \right]
\frac{\mbox{det}(H_{1})}{\mbox{det}(H_{0})} \nonumber \\ 
&&- Y_{2}(u_{j}^{(0)}) \sinh(u_{j}^{(0)}) \sinh(u_{j}^{(0)} + \gamma) \prod_{i=1}^{L-2} \frac{\sinh(u_{i}^{(2)} - u_{j}^{(0)})}{\sinh(u_{i}^{(2)})}
\frac{\sinh(u_{i}^{(2)} + \gamma + u_{j}^{(0)})}{\sinh(u_{i}^{(2)} + \gamma)}  \frac{\mbox{det}(H_{2})}{\mbox{det}(H_{0})} \nonumber \\
&& \;\;\;\;\;\;\;\;\;\;\;\;\;\;\;\;\;\;\;\;\;\;\;\;\;\;\;\;\;\;\;\;\;\;\;\;\;\;\;\;\;\;\;\;\;\;\;\;
\;\;\;\;\;\;\;\;\;\;\;\;\;\;\;\;\;\;\;\;\;\;\;\;\;\;\;\;\;\;\;\;\;\;\;\;\;\;\;\;\;\;\;\;\;\;\;\;\;\;\;\;\;\;\;\;\;\;\; j=1, \dots, L+2 \nonumber \\
\end{eqnarray}

\begin{eqnarray}
&& \left[ 2 \cosh(\gamma) \sinh(h_{1}^{+}) \sinh(h_{1}^{-}) \sinh(\gamma)^{2L}  \prod_{i=1}^{L+2} \frac{ \sinh( u_{i}^{(0)} - u_{j}^{(1)} )}{\sinh(u_{i}^{(0)})}
\frac{ \sinh( u_{i}^{(0)} +\gamma + u_{j}^{(1)} )}{\sinh(u_{i}^{(0)}+\gamma)} - Y_{0}(u_{j}^{(1)}) \right] = \nonumber \\
&&  + Y_{2}(u_{j}^{(1)}) \sinh(u_{j}^{(1)}) \sinh(u_{j}^{(1)} + \gamma) \prod_{i=1}^{L-2} \frac{\sinh(u_{i}^{(2)} - u_{j}^{(1)})}{\sinh(u_{i}^{(2)})}
\frac{\sinh(u_{i}^{(2)} + \gamma + u_{j}^{(1)})}{\sinh(u_{i}^{(2)} + \gamma)}  \frac{\mbox{det}(H_{2})}{\mbox{det}(H_{0})} \nonumber \\
&& + Y_{1}(-u_{j}^{(1)}-\gamma) \prod_{i=1}^{L-1} \frac{\sinh(u_{i}^{(1)} + \gamma + u_{j}^{(1)})}{\sinh(u_{i}^{(1)})} \frac{\mbox{det}(H_{1})}{\mbox{det}(H_{0})}
\;\;\;\;\;\;\;\;\;\;\;\;\;\;\;\;\;\;\;\;\;\;\;\;\;\;\;\;\;\;\;\;\;\;\;\;\;\;\;\;\;\;\;\; j=1, \dots, L-1 \nonumber \\
\end{eqnarray}
\begin{eqnarray}
\label{DF}
&& \left[ 2 \cosh(\gamma) \sinh(h_{1}^{+}) \sinh(h_{1}^{-}) \sinh(\gamma)^{2L}  \prod_{i=1}^{L+2} \frac{ \sinh( u_{i}^{(0)} - u_{j}^{(2)} )}{\sinh(u_{i}^{(0)})}
\frac{ \sinh( u_{i}^{(0)} +\gamma + u_{j}^{(2)} )}{\sinh(u_{i}^{(0)}+\gamma)} - Y_{0}(u_{j}^{(2)}) \right] = \nonumber \\
&&\left[ Y_{1}(u_{j}^{(2)}) \prod_{i=1}^{L-1} \frac{\sinh(u_{i}^{(1)} - u_{j}^{(2)})}{\sinh(u_{i}^{(1)})}
+ Y_{1}(-u_{j}^{(2)}-\gamma) \prod_{i=1}^{L-1} \frac{\sinh(u_{i}^{(1)} + \gamma + u_{j}^{(2)})}{\sinh(u_{i}^{(1)})} \right]
\frac{\mbox{det}(H_{1})}{\mbox{det}(H_{0})}  \nonumber \\
&& \;\;\;\;\;\;\;\;\;\;\;\;\;\;\;\;\;\;\;\;\;\;\;\;\;\;\;\;\;\;\;\;\;\;\;\;\;\;\;\;\;\;\;\;\;\;\;\;\;\;
\;\;\;\;\;\;\;\;\;\;\;\;\;\;\;\;\;\;\;\;\;\;\;\;\;\;\;\;\;\;\;\;\;\;\;\;\;\;\;\;\;\;\;\;\;\;\;\;\;\;\;\;\;\;\;\;\;\;\;\;\;\;\;\;\;\;\; j=1, \dots, L-2 \nonumber \\
\end{eqnarray}
where $H_{i}$ are $2 \times 2$ matrices resulting from the solution of the system of equations
(\ref{rts}) for the ratios $\frac{f_{1}(0)}{f_0}$ and $\frac{g_{2}}{f_0}$. These matrices turn out to be
\begin{eqnarray}
H_{0} = \pmatrix{ 
\phi_{1}(u_{1}) & \phi_{2}(u_{1}) \cr
\phi_{1}(u_{2}) & \phi_{2}(u_{2}) \cr} \;\;\;\;\;\;\;\;\;\;
H_{1} = \pmatrix{ 
\phi_{0}(u_{1}) & \phi_{2}(u_{1}) \cr
\phi_{0}(u_{2}) & \phi_{2}(u_{2}) \cr} \;\;\;\;\;\;\;\;\;\;
H_{2} = \pmatrix{ 
\phi_{1}(u_{1}) & \phi_{0}(u_{1}) \cr
\phi_{1}(u_{2}) & \phi_{0}(u_{2}) \cr} \;\;\;\;
\end{eqnarray}
where 
\begin{eqnarray}
\phi_{0}(u) &=& 2 \cosh(\gamma) \sinh(h_{1}^{+}) \sinh(h_{1}^{-}) \sinh(\gamma)^{2L}  \prod_{i=1}^{L+2} \frac{ \sinh( u_{i}^{(0)} + \frac{\gamma}{2} - u )}{\sinh(u_{i}^{(0)})}
\frac{\sinh( u_{i}^{(0)} + \frac{\gamma}{2} + u )}{\sinh(u_{i}^{(0)}+\gamma)} \nonumber \\
&-& Y_{0}(u - \frac{\gamma}{2}) \nonumber \\
\phi_{1}(u) &=& Y_{1}(-u-\frac{\gamma}{2}) \prod_{i=1}^{L-1} \frac{\sinh(u_{i}^{(1)} + \frac{\gamma}{2} + u)}{\sinh(u_{i}^{(1)})} \nonumber \\
\phi_{2}(u) &=& Y_{2}(u - \frac{\gamma}{2}) \sinh(u-\frac{\gamma}{2}) \sinh(u + \frac{\gamma}{2}) \prod_{i=1}^{L-2} \frac{\sinh(u_{i}^{(2)} + \frac{\gamma}{2} - u)}{\sinh(u_{i}^{(2)})}
\frac{\sinh(u_{i}^{(2)} + \frac{\gamma}{2} + u)}{\sinh(u_{i}^{(2)} + \gamma)}. \nonumber \\
\end{eqnarray}

Taking into account the Eq. (\ref{ft}) we
are left with the following expression for the eigenergies
$E$ of the hamiltonian (\ref{ham}),
\begin{equation}
\label{EE}
E = -\sinh(\gamma)^2 \sum_{i=1}^{L+2} \frac{1}{\sinh(u_{i}^{(0)}) \sinh(u_{i}^{(0)}+ \gamma)}
-L \cosh(\gamma) - \frac{\sinh(\gamma)^2}{\cosh(\gamma)},
\end{equation}
given in terms of the roots $u_{i}^{(0)}$.
We end this section by remarking that numerical checks performed for $L=2,3,4$ show that the spectrum generated by
the Eqs. (\ref{D1})-(\ref{DF}) is complete.

\section{Concluding Remarks}

In this paper we have proposed a functional method in the theory of exactly solvable models based
on the Yang-Baxter algebra. Using this method we were able to derive the eigenvalues of the $XXZ$ model
with non-diagonal twists and open boundaries for general values of the bulk anisotropy and boundary parameters.
Our solution is presented in terms of analogues of Bethe ansatz equations whose variables involved are
precisely the roots of the transfer matrix eigenvalues and roots of auxiliary functions defined in
terms of the monodromy matrix elements.

Concerning the $XXZ$ model with non-diagonal twists discussed in the section $2$, we stress the
unusual dependence of the Eqs. (\ref{baf}) and (\ref{baf1}) with the chain lengh $L$. We hope the
computation of physical properties in the thermodynamical limit $L \rightarrow \infty$, such as the
interfacial tension obtained in \cite{BATCH1}, will be benefited by the use of
Eqs. (\ref{baf}, \ref{baf1}). As shown in \cite{MAR}, we also remark that
our solution corresponds to the eigenvalues of Baxter's eight vertex model
with elliptic modulus $\kappa =1$ and a particular choice of diagonal twist matrix.

Regarding the case of non-diagonal open boundaries described in the section $3$, we remark the solution
recently presented for general values of the anisotropy and boundary parameters obtained from the
representation theory of $q$-Onsager algebra \cite{BASE} and the multiple reference state structure found
in \cite{ZZ1,ZZ}. An interesting problem would be unveiling the connection between these approaches. Since our
method is based on the Yang-Baxter algebra, which is a common algebraic structure underlying integrable
vertex models, we expect that our approach can be applied for other models with general open boundary 
conditions based on $q$-deformed Lie algebras \cite{LIMA}.

Finally, we observe that our solutions involve more than one kind of variable resembling the so-called
{\it nested} Bethe ansatz equations, which are typical of models with higher rank symmetry. We remark that
a similar result had been reported previously in the literature obtained  from generalized $T-Q$ relations 
\cite{NEPO3}.

\section{Acknowledgements} 
The author thanks M.J. Martins for having introduced him to these problems and for several discussions.
The author also thanks FAPESP for financial support.

\newpage
\section*{\bf Appendix A: The condition $\alpha=\beta=1$}
\setcounter{equation}{0}
\renewcommand{\theequation}{A.\arabic{equation}}

Let $\tilde{T}(\lambda)$ denote the transfer matrix
\begin{equation}
\label{transA}
\tilde{T}(\lambda) = \mbox{Tr}_{\mathcal{A}} \left[ \tilde{G}_{\mathcal{A}}
\mathcal{L}_{\mathcal{A} L}(\lambda) \mathcal{L}_{\mathcal{A} L-1}(\lambda) \dots \mathcal{L}_{\mathcal{A} 1}(\lambda) \right]
\end{equation}
where $\mathcal{L}_{\mathcal{A} j}(\lambda)$ is the $\mathcal{L}$-operator given in (\ref{loper}) and 
$\tilde{G}_{\mathcal{A}}$ is the following twist matrix
\begin{equation}
\tilde{G}_{\mathcal{A}} = \pmatrix{
0 & \alpha \cr
\beta & 0 \cr}.
\end{equation}
Now considering the matrix $\mathcal{M} = \pmatrix{
\sqrt{\alpha} & 0 \cr
0 & \sqrt{\beta} \cr}$ one can verify that $\mathcal{M}_{\mathcal{A}}^{-1} \tilde{G}_{\mathcal{A}}
\mathcal{M}_{\mathcal{A}} = \sqrt{\alpha \beta} \; G_{\mathcal{A}}$ where
$G_{\mathcal{A}}$ is given by
\begin{equation}
G_{\mathcal{A}} = \pmatrix{
0 & 1 \cr
1 & 0 \cr},
\end{equation}
which corresponds to $\tilde{G}_{\mathcal{A}}$ with $\alpha=\beta=1$. By noticing that $\mathcal{M}_{j}^{-1} \mathcal{M}_{\mathcal{A}}^{-1} 
\mathcal{L}_{\mathcal{A} j}(\lambda) \mathcal{M}_{\mathcal{A}} \mathcal{M}_{j} = \mathcal{L}_{\mathcal{A} j}(\lambda)$,
we then find the relation
\begin{equation}
\prod_{j=1}^{L} \mathcal{M}_{j}^{-1} \; \tilde{T}(\lambda) \; \prod_{i=j}^{L} \mathcal{M}_{j} = 
\sqrt{\alpha \beta} \; T(\lambda),
\end{equation} 
where $T(\lambda)$ is the transfer matrix (\ref{trii}) considered in the section $2$.
Thus the eigenvalues $\tilde{\Lambda}(\lambda)$ of the transfer matrix $\tilde{T}(\lambda)$ are
related to the ones of $T(\lambda)$ by
\begin{equation}
\tilde{\Lambda}(\lambda) = \sqrt{\alpha \beta} \; \Lambda (\lambda).
\end{equation}

\section*{\bf Appendix B: The Eigenvalues $\Lambda(0)$ and $\Delta(0)$}
\setcounter{equation}{0}
\renewcommand{\theequation}{B.\arabic{equation}}

In this appendix we consider the diagonalization of the transfer matrices
$T(0)$ and $t(0)$ associated with the $XXZ$ model with twisted and open boundary conditions
respectively.

\begin{itemize}
\item {\bf Twisted Boundary Conditions:}
\end{itemize}
The $\mathcal{L}$-operator (\ref{loper}) consist of a regular solution of the Yang-Baxter
equation, i.e. $\mathcal{L}(0) = \sinh(\gamma) P$, where $P$ denotes the permutation operator.
In this way the transfer matrix (\ref{trii}) at the point $\lambda=0$ can be written as
$T(0)=\sinh(\gamma)^{L} \hat{O}$ where the operator $\hat{O}$ is given by
\begin{equation}
\hat{O}=\mbox{Tr}_{\mathcal{A}} \left[ G_{\mathcal{A}} P_{\mathcal{A} L} P_{\mathcal{A} L-1} \dots P_{\mathcal{A} 1} \right].
\end{equation}
Using the permutator algebra, $\displaystyle P^{2}=\mbox{Id} \otimes \mbox{Id}$ and $P_{\mathcal{A} j} P_{\mathcal{A} i}= P_{\mathcal{A} i} P_{ij}$,
we can readily obtain the following expression for $\hat{O}$,
\begin{equation}
\hat{O} = G_{1} P_{1 L} \dots P_{13} P_{12},
\end{equation}
and its $n$ times product
\begin{equation}
\label{on}
\hat{O}^{n} = G_{1} G_{2} \dots G_{n} \prod_{j=1}^{n} P_{j L} P_{j L-1} \dots P_{j n+1} .
\end{equation}
From the Eq. (\ref{on}) we have
\begin{equation}
\hat{O}^{L} =  G_{1} G_{2} \dots G_{L},
\end{equation}
which implies, with $G=\pmatrix{
0 & 1 \cr
1 & 0 \cr}$, in $\hat{O}^{2L} = I$ where $I$ is the identity matrix. Thus the eigenvalues $O$
of the operator $\hat{O}$ are given by
\begin{equation}
O = e^{\frac{i\pi r}{L}} \;\;\;\;\;\;\;\;\;\;\;\;\;\;\;\;\;\;\;\;\;\;\;\;\;\;\;\;\;\;\;\;\; r = 0,1,\dots, 2L-1 \;
\end{equation}
leaving us with the following expression for $\Lambda(0)$,
\begin{equation}
\Lambda(0) = \sinh(\gamma)^{L} e^{\frac{i\pi r}{L}} \;\;\;\;\;\;\;\;\;\;\;\;\;\;\;\;\; r = 0,1,\dots, 2L-1 \; .
\end{equation}

\begin{itemize}
\item {\bf Open Boundary Conditions:}
\end{itemize}
The matrix $K^{-}_{\mathcal{A}}(u)$ presented in Eqs. (\ref{KMat}) and (\ref{kpw}) consist
of a regular solution of the reflection equation, i.e. $K^{-}_{\mathcal{A}}(0)=\sinh(h_{1}^{-}) \mbox{Id}$.
Together with the regularity property of the $\mathcal{L}$-operator (\ref{loper}) and the
permutator property $\displaystyle P^{2} = \mbox{Id} \otimes \mbox{Id}$, the double-row transfer
matrix (\ref{dtr}) at the point $u=0$ is given by
\begin{equation}
t(0) = \sinh(\gamma)^{2L} \sinh(h_{1}^{-}) \mbox{Tr}_{\mathcal{A}} \left[ K^{+}_{\mathcal{A}}(0) \right] I.
\end{equation}
Therefore, we can see that the transfer matrix $t(0)$ is proportional to the identity matrix $I$ with the following
eigenvalue
\begin{equation}
\Delta(0) = 2 \cosh(\gamma) \sinh(h_{1}^{+}) \sinh(h_{1}^{-}) \sinh(\gamma)^{2L}.
\end{equation}

\section*{\bf Appendix C: The functions $F_{i}(\lambda)$ and $f_{i}(\lambda)$}
\setcounter{equation}{0}
\renewcommand{\theequation}{C.\arabic{equation}}

In this appendix we derive the form of the functions $F_{i}(\lambda)$ and $f_{i}(\lambda)$
used in the sections $2$ and $3$, considering the algebraic properties of the
monodromy matrices $\mathcal{T}_{\mathcal{A}}(\lambda)$ and
$\bar{\mathcal{T}}_{\mathcal{A}}(\lambda)$.

In order to determine the dependence of the functions $F_{i}(\lambda)$ with the spectral parameter
$\lambda$, let us first introduce the following notation for the monodromy matrix (\ref{mono})
\begin{equation}
\label{monoL}
\mathcal{T}^{(L)}_{\mathcal{A}}(\lambda) =\pmatrix{
A_{L}(\lambda) & B_{L}(\lambda) \cr
C_{L}(\lambda) & D_{L}(\lambda) \cr},
\end{equation}
differing from the one used in (\ref{monorep}) by the index $L$ inserted
to emphasize we are considering the ordered product of $L$ matrices $\mathcal{L}_{\mathcal{A} j}(\lambda)$.
For instance, the matrices
$\mathcal{L}_{\mathcal{A} j}(\lambda)$ are given by
\begin{equation}
\label{LAJ}
\mathcal{L}_{\mathcal{A} j}(\lambda)=\pmatrix{
\alpha_{j}(\lambda) & \beta_{j}(\lambda) \cr
\gamma_{j}(\lambda) & \delta_{j}(\lambda) \cr}
\end{equation}
where
\begin{eqnarray}
\label{abcd}
\alpha_{j}(\lambda)&=&\pmatrix{
a(\lambda) & 0 \cr
0 & b(\lambda) \cr}_{j} \;\;\;\;\;\;\;\;\;\;\;\;\;\;\;\;\;\;\;\;\;
\beta_{j}(\lambda)=\pmatrix{
0 & 0 \cr
c(\lambda) & 0 \cr}_{j} \nonumber \\
\gamma_{j}(\lambda)&=&\pmatrix{
0 & c(\lambda) \cr
0 & 0 \cr}_{j} \;\;\;\;\;\;\;\;\;\;\;\;\;\;\;\;\;\;\;\;\;\;\;\;\;\;
\delta_{j}(\lambda)=\pmatrix{
b(\lambda) & 0 \cr
0 & a(\lambda) \cr}_{j}
\end{eqnarray}
are matrices acting non-trivially in the $j$-th site of the chain.

Moreover, the monodromy matrix $\mathcal{T}^{(L)}_{\mathcal{A}}(\lambda)$
defined by Eq. (\ref{mono}) satisfies the recurrence relation
\begin{equation}
\label{REC}
\mathcal{T}^{(L+1)}_{\mathcal{A}}(\lambda) = \mathcal{L}_{\mathcal{A} L+1}(\lambda)
\mathcal{T}^{(L)}_{\mathcal{A}}(\lambda),
\end{equation}
corresponding to the following relations for its matrix elements
\begin{eqnarray}
\label{REC1}
A_{L+1}(\lambda) &=& \alpha_{L+1}(\lambda) A_{L}(\lambda) + \beta_{L+1}(\lambda) C_{L}(\lambda) \nonumber \\
B_{L+1}(\lambda) &=& \alpha_{L+1}(\lambda) B_{L}(\lambda) + \beta_{L+1}(\lambda) D_{L}(\lambda) \nonumber \\
C_{L+1}(\lambda) &=& \gamma_{L+1}(\lambda) A_{L}(\lambda) + \delta_{L+1}(\lambda) C_{L}(\lambda) \nonumber \\
D_{L+1}(\lambda) &=& \gamma_{L+1}(\lambda) B_{L}(\lambda) + \delta_{L+1}(\lambda) D_{L}(\lambda).
\end{eqnarray}

At this stage, it is important to keep in mind that the spectral parameter $\lambda$ enters in the monodromy
matrix only through the Boltzmann weights $a(\lambda)$ and $b(\lambda)$, since $c(\lambda)$ is
$\lambda$ independent (\ref{loper}). For the forthcoming discussion, it is convenient to introduce
the weight $w(\lambda)$ only to denote the Boltzmann weights possessing dependence with  $\lambda$.
Thus, $w(\lambda)$ can represent both $a(\lambda)$ and $b(\lambda)$. The reason for the introduction of $w(\lambda)$
will become clear in what follows. From the Eqs. (\ref{mono}), (\ref{LAJ}) and (\ref{abcd}) it is clear
that $B_{1}(\lambda)$ and $C_{1}(\lambda)$ are polynomials of degree $0$ in $w$, while
$A_{1}(\lambda)$ and $D_{1}(\lambda)$ are polynomials of degree $1$ in $w$.
These initial conditions allow us to use the Eqs. (\ref{REC1}) to determine the degree of the
elements of $\mathcal{T}_{\mathcal{A}}^{(L)}(\lambda)$ in $w$ for arbitrary lengh $L$. Thus we obtain that
$B_{L}(\lambda)$ and $C_{L}(\lambda)$ are polynomials of degree $L-1$ in $w$, while the operators
$A_{L}(\lambda)$ and $D_{L}(\lambda)$ are of degree $L$.

In terms of the variables $x=e^{\lambda}$ and $q=e^{\gamma}$ the Boltzmann weights $a(\lambda)$ and
$b(\lambda)$ can be written as $a(\lambda)=\frac{1}{2}\left( xq - x^{-1} q^{-1} \right)$ and
$b(\lambda)=\frac{1}{2}\left( x - x^{-1} \right)$. Thus $w(\lambda)$ is of the form
$w(\lambda) \sim x + x^{-1}$, and expanding its powers we find
\begin{eqnarray}
B(\lambda) &\sim&  x^{L-1} + \dots + x^{-(L-1)} \nonumber \\
&=& x^{-(L-1)} \left[ c_{L-1} (x^2)^{L-1} + c_{L} (x^2)^{L} + \dots + c_{0} \right] .
\end{eqnarray}
Now considering the functions $F_{1}(\lambda) = \bra{\psi} B(\lambda) \ket{0}$ and
$F_{2}(\lambda) = \bra{\psi} B(\lambda) B(\lambda) \ket{0}$ used in section $2$, and the fact
that $\ket{\psi}$ is $\lambda$ independent, we can conclude that
\begin{eqnarray}
F_{1}(\lambda) &=& x^{-(L-1)} P_{1}(x) \nonumber \\
F_{2}(\lambda) &=& x^{-2(L-1)} P_{2}(x),
\end{eqnarray}
where $P_{1}(x)$ and $P_{2}(x)$ are polynomials in the variable $x^2$ of degrees $L-1$ and $2(L-1)$ respectively.
Therefore, the polynomials $P_{1}(x)$ and $P_{2}(x)$ can be written as
\begin{eqnarray}
P_{1}(x) &=& p_{1} \prod_{i=1}^{L-1} \left[ x^2 - (x_{i}^{(1)})^2 \right] \nonumber \\
P_{2}(x) &=& p_{2} \prod_{i=1}^{2(L-1)} \left[ x^2 - (x_{i}^{(2)})^2 \right]
\end{eqnarray}
where $p_{j}$ are constants and $x_{i}^{(j)}$ denote the zeroes of the polynomial $P_{j}(x)$.
Using the variables $x_{i}^{(j)}=e^{\lambda_{i}^{(j)}}$ we then have
\begin{eqnarray}
F_{1}(\lambda) &=& F_{1}(0) \prod_{i=1}^{L-1} \frac{\sinh(\lambda_{i}^{(1)} - \lambda)}{\sinh(\lambda_{i}^{(1)})} \nonumber \\
F_{2}(\lambda) &=& F_{2}(0) \prod_{i=1}^{2(L-1)} \frac{\sinh(\lambda_{i}^{(2)} - \lambda)}{\sinh(\lambda_{i}^{(2)})}
\end{eqnarray}
where the constant $p_{j}$ is absorved by $F_{j}(0)$.

Let us now consider the monodromy matrix $\bar{\mathcal{T}}_{\mathcal{A}}^{(L)}(\lambda) = \mathcal{L}_{\mathcal{A} 1}(\lambda)
\mathcal{L}_{\mathcal{A} 2}(\lambda) \dots \mathcal{L}_{\mathcal{A} L}(\lambda)$, with matrix elements
\begin{equation}
\bar{\mathcal{T}}_{\mathcal{A}}^{(L)}(\lambda)=\pmatrix{
\bar{A}_{L}(\lambda) & \bar{B}_{L}(\lambda) \cr
\bar{C}_{L}(\lambda) & \bar{D}_{L}(\lambda) \cr},
\end{equation}
appearing in the construction of the double-row transfer matrix (\ref{dtr}). Compared to the notation
previously used (\ref{monobar}), we have also included the index $L$ stressing we are considering the ordered product of $L$ matrices
$\mathcal{L}_{\mathcal{A} j}(\lambda)$. The monodromy matrix $\bar{\mathcal{T}}_{\mathcal{A}}^{(L)}(\lambda)$
obeys the recurrence formula
\begin{equation}
\bar{\mathcal{T}}_{\mathcal{A}}^{(L+1)}(\lambda) =
\bar{\mathcal{T}}_{\mathcal{A}}^{(L)}(\lambda) \mathcal{L}_{\mathcal{A} L+1}(\lambda),
\end{equation}
corresponding to the following recursion relations for the matrix elements
\begin{eqnarray}
\label{REC2}
\bar{A}_{L+1}(\lambda) &=& \bar{A}_{L}(\lambda) \alpha_{L+1}(\lambda) + \bar{B}_{L}(\lambda) \gamma_{L+1}(\lambda) \nonumber \\
\bar{B}_{L+1}(\lambda) &=& \bar{A}_{L}(\lambda) \beta_{L+1}(\lambda) + \bar{B}_{L}(\lambda) \delta_{L+1}(\lambda) \nonumber \\
\bar{C}_{L+1}(\lambda) &=& \bar{C}_{L}(\lambda) \alpha_{L+1}(\lambda) + \bar{D}_{L}(\lambda) \gamma_{L+1}(\lambda) \nonumber \\
\bar{D}_{L+1}(\lambda) &=& \bar{C}_{L}(\lambda) \beta_{L+1}(\lambda) + \bar{D}_{L}(\lambda) \delta_{L+1}(\lambda).
\end{eqnarray}

The same arguments previously used for the elements of $\mathcal{T}_{\mathcal{A}}^{(L)}(\lambda)$ can
now be considered for the elements of $\bar{\mathcal{T}}_{\mathcal{A}}^{(L)}(\lambda)$.
Thus we obtain that the operators $\bar{B}_{L}(\lambda)$ and
$\bar{C}_{L}(\lambda)$ are polynomials of degree $L-1$ in $w$ while
$\bar{A}_{L}(\lambda)$ and $\bar{D}_{L}(\lambda)$ are of degree $L$. Hence the function
$f_{1}(u) = \bra{\Phi} B(u) \ket{0}$ required in the section 3 can be written as
\begin{equation}
\label{ww1}
f_{1}(u) = f_{1}(0) \prod_{i=1}^{L-1} \frac{\sinh(u_{i}^{(1)} - u)}{\sinh(u_{i}^{(1)})}.
\end{equation} 
On the other hand the function $f_{2}(u) = \bra{\Phi} B(u) \bar{B}(u) \ket{0}$ requires an extra
analysis. Noticing that $\alpha_{j}(0) = \frac{1}{c(0)}\gamma_{j}(0) \beta_{j}(0)$ and
$\delta_{j}(0) = \frac{1}{c(0)}\beta_{j}(0) \gamma_{j}(0)$ together with
$\beta_{j}^{2}(\lambda) = \gamma_{j}^{2}(\lambda)=0$, one can easily see from the Eqs. (\ref{REC1})
and (\ref{REC2}) that $B(0)\bar{B}(0)=0$. This same analysis can be performed for the point
$\lambda = -\gamma$ yielding $B(-\gamma)\bar{B}(-\gamma)=0$. Thus the function $f_{2}(u)$
can be written as
\begin{equation}
\label{ww2}
f_{2}(u) = g_{2} \sinh(u) \sinh(u+\gamma) \prod_{i=1}^{2(L-2)} \frac{\sinh(u_{i}^{(2)} - u)}{\sinh(u_{i}^{(2)})}.
\end{equation}
In the section $3$ we also require the function $\Delta(u)$ defined as
\begin{equation}
\Delta(u) = \frac{\bra{\Phi} t(u) \ket{\Phi}}{\left\langle \Phi | \Phi \right\rangle}.
\end{equation}
The above results together with the Eqs. (\ref{dbr}), (\ref{UU}), and (\ref{kpw})
show that the double-row transfer matrix consist of a polynomial of degree $2(L+2)$ in $w$. Thus the
eigenvalues $\Delta(u)$ are of the form
\begin{equation}
\label{w3}
\Delta(u) = \Delta(0) \prod_{i=1}^{2(L+2)} \frac{\sinh(u_{i}^{(0)} - u )}{\sinh(u_{i}^{(0)})},
\end{equation}
where $u_{i}^{(0)}$ denote the zeroes of $\Delta(u)$.

\section*{\bf Appendix D: Crossing symmetry}
\setcounter{equation}{0}
\renewcommand{\theequation}{D.\arabic{equation}}

In this appendix we deduce some important properties used in the sections $2$ and $3$
which are consequence of the crossing symmetry.

Besides the Yang-Baxter equation, the $\mathcal{L}$-operator (\ref{loper})
satisfies other important relations, namely
\begin{eqnarray}
\label{TS}
&&\mbox{Temporal invariance:} \;\;\;\;\;\;\;\;\;\;\; \mathcal{L}_{12}^{t_{1} t_{2}} (\lambda) = \mathcal{L}_{12} (\lambda) \nonumber \\
\label{CS}
&&\mbox{Crossing symmetry:} \;\;\;\;\;\;\;\;\;\;\;\;\; \mathcal{L}_{12}(\lambda) =
- V_{1} \mathcal{L}_{12}^{t_{2}}(-\lambda-\gamma) V_{1}^{-1}
\end{eqnarray}
where $V=\pmatrix{
0 & 1 \cr
-1 & 0 \cr}$ and $t_{i}$ stands for the transposition in the space with index $i$.

In order to estabilish a relation between the matrices $\mathcal{T}_{\mathcal{A}}(\lambda)$
and $\bar{\mathcal{T}}_{\mathcal{A}}(\lambda)$, we first recall the definitions
\begin{eqnarray}
\mathcal{T}_{\mathcal{A}}(\lambda) &=& \mathcal{L}_{\mathcal{A} L}(\lambda)
\mathcal{L}_{\mathcal{A} L-1}(\lambda) \dots \mathcal{L}_{\mathcal{A} 1}(\lambda) \\
\bar{\mathcal{T}}_{\mathcal{A}}(\lambda) &=& \mathcal{L}_{\mathcal{A} 1}(\lambda)
\dots \mathcal{L}_{\mathcal{A} L-1}(\lambda) \mathcal{L}_{\mathcal{A} L}(\lambda) ,
\end{eqnarray}
and observe that $\mathcal{L}_{12}^{t_{1}} (\lambda) = \mathcal{L}_{12}^{t_{2}} (\lambda)$
which follows from the temporal invariance (\ref{TS}).
Next we consider the transposition in the auxiliary space $\mathcal{A}$,
\begin{eqnarray}
\mathcal{T}^{t_{\mathcal{A}}}_{\mathcal{A}}(\lambda)&=&
\mathcal{L}^{t_{\mathcal{A}}}_{\mathcal{A} 1}(\lambda) \dots \mathcal{L}^{t_{\mathcal{A}}}_{\mathcal{A} L}(\lambda) \nonumber \\
&=& \mathcal{L}^{t_{1}}_{\mathcal{A} 1}(\lambda) \dots \mathcal{L}^{t_{L}}_{\mathcal{A} L}(\lambda),
\end{eqnarray}
which allows us to use the crossing symmetry (\ref{CS}) to obtain
\begin{eqnarray}
\label{B1}
\mathcal{T}^{t_{\mathcal{A}}}_{\mathcal{A}}(\lambda)&=&
(-1)^{L} V_{\mathcal{A}}^{-1} \mathcal{L}_{\mathcal{A} 1}(-\lambda-\gamma) V_{\mathcal{A}} \dots
V_{\mathcal{A}}^{-1} \mathcal{L}_{\mathcal{A} L}(-\lambda-\gamma) V_{\mathcal{A}} \nonumber \\
&=& (-1)^{L} V_{\mathcal{A}}^{-1} \bar{\mathcal{T}}_{\mathcal{A}}(-\lambda-\gamma) V_{\mathcal{A}}.
\end{eqnarray}

Now we also consider the transposition $t=t_{1} \dots t_{L}$. Then we find
\begin{eqnarray}
\mathcal{T}^{t_{\mathcal{A}} t}_{\mathcal{A}}(\lambda)&=&
\left[ \mathcal{L}_{\mathcal{A} L}(\lambda) \dots \mathcal{L}_{\mathcal{A} 1}(\lambda) \right]^{t_{\mathcal{A}} t_{1} \dots t_{L}} \nonumber \\
&=& \mathcal{L}^{t_{\mathcal{A}} t_{1}}_{\mathcal{A} 1}(\lambda) \dots \mathcal{L}^{t_{\mathcal{A}} t_{L}}_{\mathcal{A} L}(\lambda) \nonumber \\
&=& \mathcal{L}_{\mathcal{A} 1}(\lambda) \dots \mathcal{L}_{\mathcal{A} L}(\lambda),
\end{eqnarray}
which leads immediately to the relation
\begin{equation}
\label{B2}
\mathcal{T}^{t_{\mathcal{A}} t}_{\mathcal{A}}(\lambda)= \bar{\mathcal{T}}_{\mathcal{A}}(\lambda).
\end{equation}
In terms of the matrix elements, the relations (\ref{B1}) and (\ref{B2}) read,
\begin{eqnarray}
\label{CCR}
A(\lambda) &=& (-1)^{L} \bar{D}(-\lambda - \gamma) \;\;\;\;\;\;\;\;\;\;\;\;\;\;\;\;\;\; C(\lambda) = (-1)^{L+1} \bar{C}(-\lambda - \gamma) \nonumber \\
B(\lambda) &=& (-1)^{L+1} \bar{B}(-\lambda - \gamma) \;\;\;\;\;\;\;\;\;\;\;\;\;\;\; D(\lambda) = (-1)^{L} \bar{A}(-\lambda - \gamma) \nonumber \\
A^{t}(\lambda) &=& \bar{A}(\lambda) \;\;\;\;\;\;\;\;\;\;\;\;\;\;\;\;\;\;\;\;\;\;\;\;\;\;\;\;\;\;\;\;\;\;\;\; C^{t}(\lambda) = \bar{B}(\lambda) \nonumber \\
B^{t}(\lambda) &=& \bar{C}(\lambda) \;\;\;\;\;\;\;\;\;\;\;\;\;\;\;\;\;\;\;\;\;\;\;\;\;\;\;\;\;\;\;\;\;\;\;\; D^{t}(\lambda) = \bar{D}(\lambda).
\end{eqnarray}

A straighforward calculation taking into account the relations (\ref{CCR}) reveals that the twisted
transfer matrix (\ref{trii}) satisfies the relation
\begin{equation}
\label{DD}
T^{t}(\lambda) = (-1)^{L+1} T(-\lambda-\gamma),
\end{equation}
while we have the following identity for the double-row transfer matrix (\ref{dtr})
\begin{equation}
t(u) = t(-u-\gamma).
\end{equation}

\end{document}